\documentclass[aps,prc,twocolumn,superscriptaddress,preprintnumbers,showpacs,nofootinbib,a4paper,amsmath,amssymb]{revtex4-1}

\usepackage[latin1]{inputenc}
\usepackage{graphicx}% Include figure files
\usepackage{dcolumn}% Align table columns on decimal point
\usepackage{bm}% bold math
\usepackage{color}%
\usepackage[bookmarks=false, pdfborder={0 0 0}, colorlinks=true, linkcolor=blue, citecolor=blue, filecolor=blue, urlcolor=blue]{hyperref}

\graphicspath{{.}}
\newcommand{\hzdr}{\affiliation{Helmholtz-Zentrum Dresden-Rossendorf (HZDR), Bautzner Landstr. 400, D-01328 Dresden, Germany}}
\newcommand{\tu}{\affiliation{Technische Universit\"at Dresden, Institut f\"ur Kern- und Teilchenphysik, D-01062 Dresden, Germany}}
\newcommand{\infnpd}{\affiliation{Istituto Nazionale di Fisica Nucleare (INFN), Sezione di Padova, I-35131 Padova, Italy}}
\newcommand{\unige}{\affiliation{Dipartimento di Fisica, Universit\`a di Genova, and Istituto Nazionale di Fisica Nucleare (INFN), Sezione di Genova, Genova, Italy}}
\newcommand{\unipd}{\affiliation{Dipartimento di Fisica e Astronomia, Universit\`a di Padova, Padova, Italy}}
\newcommand{\zarqa}{\affiliation{Physics Department Hashemite University, Zarqa 13133, Jordan}}
\newcommand{\atomki}{\affiliation{MTA ATOMKI, Debrecen, Hungary}}

\begin{document}

\title{Strengths of the resonances at 436, 479, 639, 661, and 1279\,keV in the $^{22}$Ne(p,$\gamma$)$^{23}$Na reaction}

\author{Rosanna Depalo}\infnpd\unipd 			
\author{Francesca Cavanna}\unige 			
\author{Federico Ferraro}\unige 			
\author{Alessandra Slemer}\infnpd\unipd 			
\author{Tariq Al-Abdullah}\hzdr\zarqa 			
\author{Shavkat Akhmadaliev}\hzdr 			
\author{Michael Anders}\hzdr\tu 			
\author{Daniel Bemmerer}\email{d.bemmerer@hzdr.de}\hzdr	
\author{Zolt\'an Elekes}\hzdr\atomki				
\author{Giovanni Mattei}\unipd 			
\author{Stefan Reinicke}\hzdr\tu 			
\author{Konrad Schmidt}\hzdr\tu 			
\author{Carlo Scian}\unipd 			
\author{Louis Wagner}\hzdr\tu 				

\date{\today}

\begin{abstract}
The $^{22}$Ne(p,$\gamma$)$^{23}$Na reaction is included in the neon-sodium cycle of hydrogen burning. A number of narrow resonances in the Gamow window dominate the thermonuclear reaction rate. Several resonance strengths are only poorly known. As a result, the $^{22}$Ne(p,$\gamma$)$^{23}$Na thermonuclear reaction rate is the most uncertain rate of the cycle.
Here, a new experimental study of the strengths of the resonances at 436, 479, 639, 661, and 1279\,keV proton beam energy is reported. The data have been obtained using a tantalum target implanted with $^{22}$Ne. The strengths $\omega\gamma$ of the resonances at 436, 639, and 661\,keV have been determined with a relative approach, using the 479- and 1279-keV resonances for normalization. Subsequently, the ratio of resonance strengths of the 479- and 1279-keV resonances was determined, improving the precision of these two standards.
The new data are consistent with, but more precise than, the literature with the exception of the resonance at 661\,keV, which is found to be less intense by one order of magnitude. 
In addition, improved branching ratios have been determined for the gamma decay of the resonances at 436, 479, and 639 keV.

\end{abstract}

% PACS 2010
% 25.40.Lw	Radiative capture
% 25.40.Ny	Resonance reactions
% 26.30.-k	nucleosynthesis in novae and supernovae, 
%
% 81.70.Jb	Chemical composition analysis, chemical depth and dopant profiling
% 25.40.Ep	Inelastic proton scattering
% 61.72.S- Impurities in crystals

\maketitle

\section{\label{sec:intro} Introduction}
The neon-sodium cycle of hydrogen burning (NeNa cycle) plays a crucial role for the synthesis of the elements between $^{20}$Ne and $^{24}$Mg \cite{Marion57-ApJ}. Whithin the cycle, $^{22}$Ne can only be produced by the decay of radioactive $^{22}$Na ($T_{1/2}$ = 2.6027 y \cite{Firestone05-NDS}). However, this decay competes with proton capture on $^{22}$Na \cite{Sallaska10-PRL}. At temperatures $T$ $>$ 70 MK, proton capture surpasses the decay, and $^{22}$Ne is effectively bypassed in the NeNa cycle. In this scenario, the destruction of $^{22}$Ne by the $^{22}$Ne(p,$\gamma$)$^{23}$Na reaction is no longer compensated by fresh production.

However, in a helium burning scenario $^{22}$Ne may be produced from the ashes of the carbon-nitrogen-oxygen (CNO) cycle, using the chain: $^{14}$N($\alpha$,$\gamma$)$^{18}$F($\beta^+ \nu$)$^{18}$O($\alpha$,$\gamma$)$^{22}$Ne. Therefore, $^{22}$Ne can still be present in second generation stars or whenever material processed by the CNO cycle is mixed with helium-rich material \cite{Clayton03-Book}. The neutrons stored in $^{22}$Ne have recently been suggested to play a role in neutron capture nucleosynthesis in supernova of type Ia that may lay the seed for an alternative production of p-nuclei \cite{Travaglio11-ApJ,Kusakabe11-ApJ,Travaglio15-ApJ}.

The $^{22}$Ne(p,$\gamma$)$^{23}$Na reaction converts neon to sodium. For the case of classical nova nucleosynthesis, a sensitivity study has shown how the uncertainty on the $^{22}$Ne(p,$\gamma$)$^{23}$Na reaction rate propagates to the abundance of the elements between $^{20}$Ne and $^{27}$Al \cite{Iliadis02-ApJSS}. For type Ia supernovae, the synthesis of $^{18}$O and $^{23,24}$Na may be affected by proton capture on $^{22}$Ne \cite{Parikh13-AA}.

The $^{22}$Ne(p,$\gamma$)$^{23}$Na excitation function is characterized by the contribution of many narrow resonances. At low energies $E_{\rm p} \leq$ 400\,keV, there is a large number of resonances that have never been observed experimentally, leading to a significant uncertainty on the reaction rate \cite{Cavanna14-EPJA}. Some of these low-energy resonances have recently been studied at the LUNA 400-kV accelerator in Gran Sasso, Italy \cite{Cavanna15-PRL}. 

The strengths of the resonances $E_{\rm p}>$ 400 keV have been measured mostly with implanted $^{22}$Ne solid targets \cite[and references therein]{Endt90-NPA,Kachan06-Izv}. Two resonances deserve special mention, as they are used as reference standards here: The 479-keV resonance strength has recently been determined relative to a well-known $^{27}$Al(p,$\gamma$)$^{28}$Si resonance \cite{Longland10-PRC, Kelly15-PRC} in an aluminum target implanted with $^{22}$Ne. The strength of the 1279-keV resonance has been measured in an absolute manner for a carbon target implanted with $^{22}$Ne, with the stoichiometric ratio derived from a Rutherford backscattering analysis \cite{Keinonen77-PRC}. 

\begin{figure}[!h]
\includegraphics[width=0.8\columnwidth]{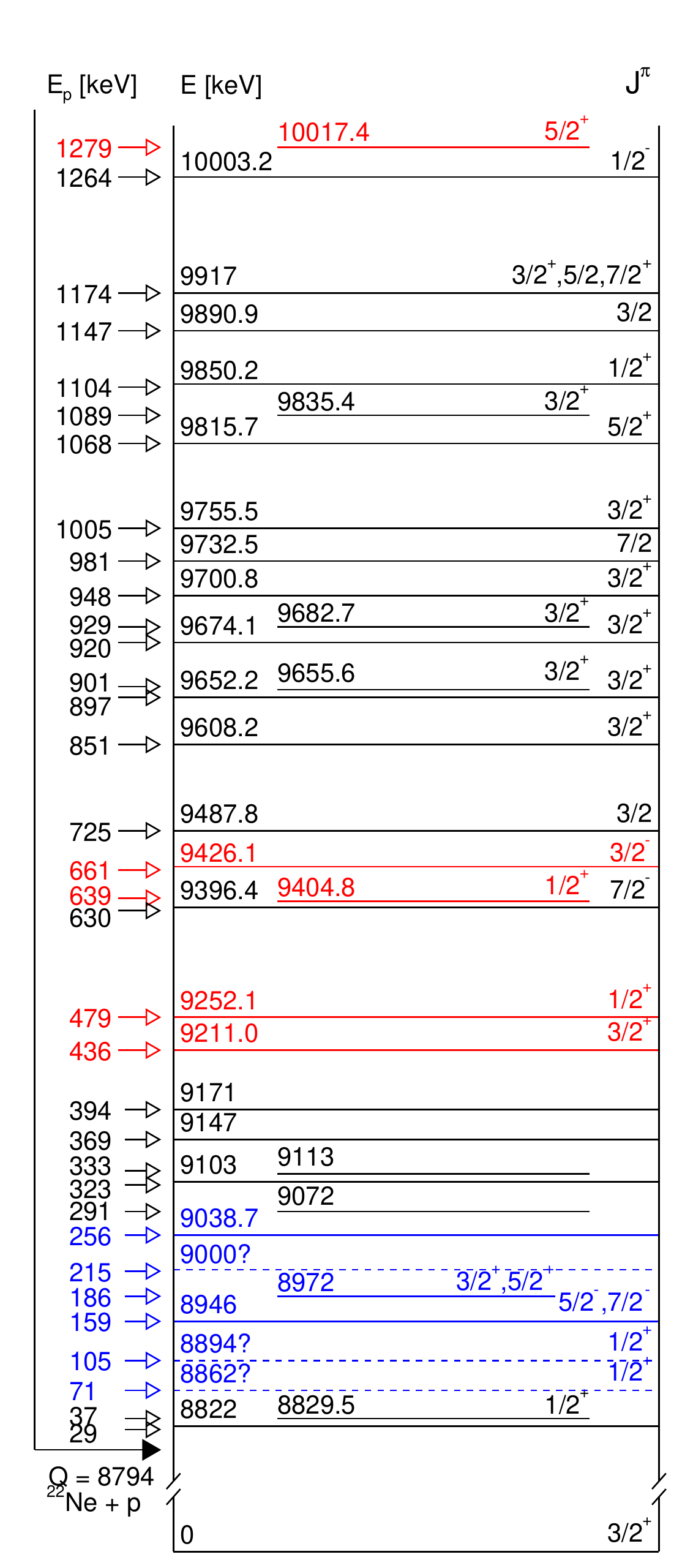}
\caption{\label{fig:ls}Partial level scheme of $^{23}$Na \cite{Firestone07-NDS23}. The excited states and corresponding resonance energies $E_{\rm p}$ investigated in the present experiment are marked in red. The resonances investigated in Ref.~\cite{Cavanna15-PRL} are marked in blue.}
\end{figure}

Here, a new experiment on the $^{22}$Ne(p,$\gamma$)$^{23}$Na resonances at 436-, 479-, 639-, 661-, and 1279-keV proton beam energy (fig. \ref{fig:ls}) is reported. These resonances contribute significantly to the reaction rate at temperatures between 0.3 and 2 GK, but their strenghts are only known with uncertainties between 10$\%$ and 30$\%$ (tab. \ref{tab:strengths}). Of the five resonances studied here, the 479-keV and 1279-keV resonances have been used as references for normalization. The combined use of two reference resonances allowed to reduce the uncertainty on the resonance strengths to 8\%. Subsequently, also the two standards have been linked to each other, improving their precision.

The present work is organized as follows: section~\ref{sec:setup} describes the experimental setup, sec.~\ref{sec:Results} includes the data analysis and results. The discussion and a summary are given in secs.~\ref{sec:Discussion} and \ref{sec:Summary}, respectively. More details can be found in thesis works \cite{Depalo15-PhD,Cavanna15-PhD,Ferraro13-Master,Slemer13-Master}.

%%%%%%%%%%%%%%%%%%%%%%%%%%%%%%%%%%%%%%%%%%%%%%
\section{\label{sec:setup} Experimental setup}

\subsection{Ion beam and target chamber}

\begin{figure}[tb]
\includegraphics[width=\columnwidth]{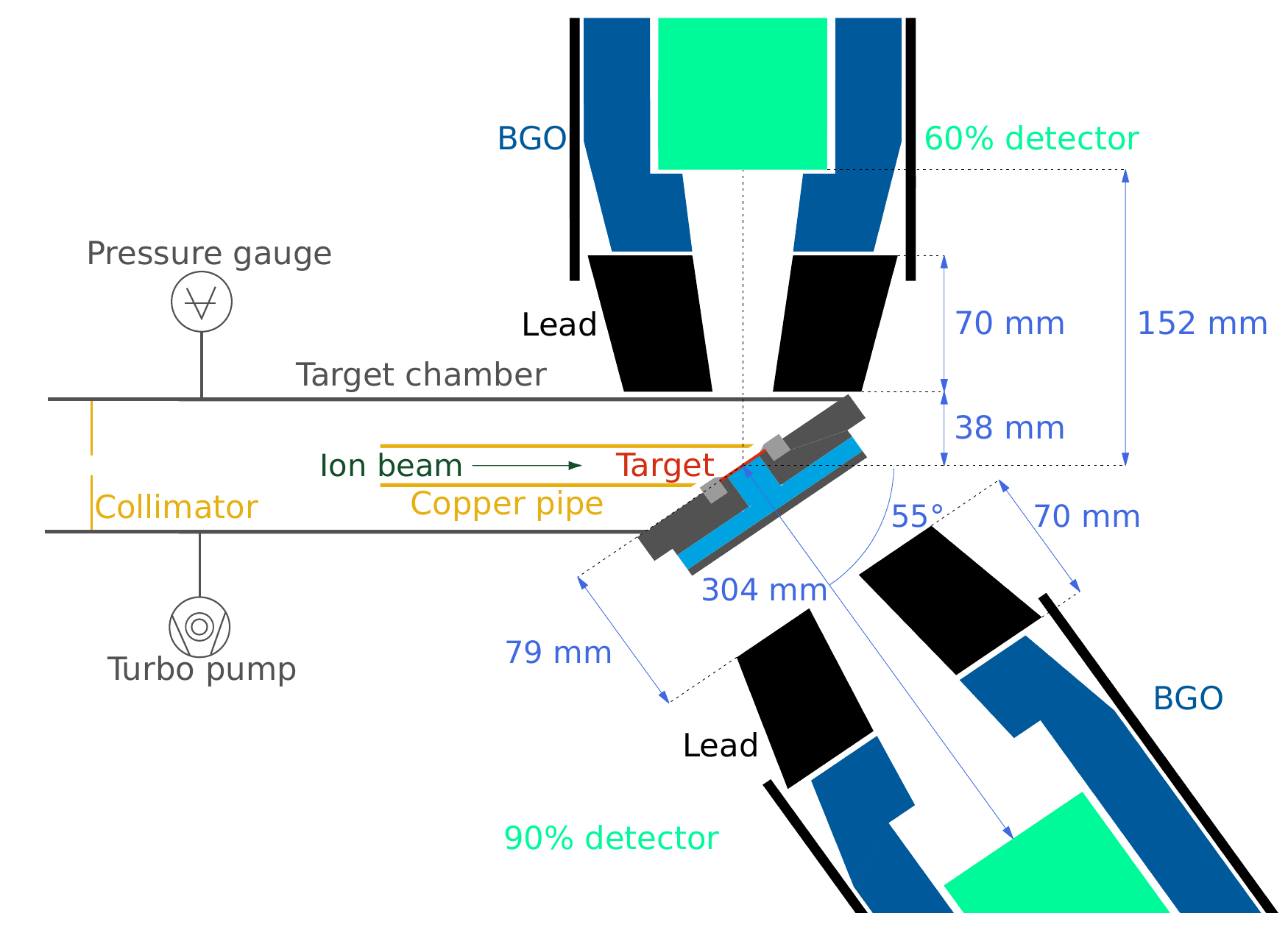}
\caption{\label{fig:setup}Schematic view of the experimental setup.}
\end{figure}

The experiment was performed at the 3-MV Tandetron accelerator of Helmholtz-Zentrum Dresden-Rossendorf. 
A proton beam of $E_{\rm p}$ = 0.4 - 1.3 MeV with intensity between 5 and 10 $\mu$A was delivered to a solid target implanted with  $^{22}$Ne (see below, sec.\,\ref{subsec:Targets}).
Before reaching the target, the beam went through a water cooled copper collimator of 5 mm diameter and a copper tube of 30 mm diameter extending up to 2 mm distance from the target (fig. \ref{fig:setup}). During irradiation, this tube was biased to $-100$\,V in order to force secondary electrons back to the target. This allowed a precise measurement of the charge collected on target. The integrated charge was measured with a current integrator and recorded both by an analog scaler and a list mode digital data acquisition system. 
The target holder was tilted at $55^{\circ}$ with respect to the beam direction. The target backing was directly water cooled to prevent deterioration. A turbomolecular pump kept the target chamber pressure in the $10^{-7}$ mbar range.

\subsection{Targets}\label{subsec:Targets}

$^{22}$Ne implanted targets were prepared at the 200-kV Danfysik 1090 ion implanter of Legnaro National Laboratories, Italy. 
Neon gas with natural isotopic composition (90.48$\%$ $^{20}$Ne, 0.27$\%$ $^{21}$Ne, 9.25$\%$ $^{22}$Ne) was ionized, accelerated and sent to an analysing magnet to select the mass-22 isotope. Then the $^{22}$Ne beam was sent to a quadrupole magnet providing beam focusing, and hence to the target chamber. A beam scanning magnet located between the quadrupole and the target produced a uniform irradiated area of (10$\times$8) cm$^2$ at the target level. This allowed to produce, in a single irradiation, more than one implanted target with uniform implantation along the backing surface.
Previous investigations on the implantation of neon gas showed that tantalum backings provide high saturation concentrations (between 10$^{17}$ and 10$^{18}$ atoms/cm$^2$) and a good stability of the implantation against beam irradiation \cite{Selin67-NIM, Seuthe87-NIMA, Giesen93-NPA, Lee09-NIMB}.  

For the present experiment, tantalum disks of 27 mm diameter and 0.22 mm thickness were subsequently implanted with the following $^{22}$Ne fluences and energies: $1.5 \cdot 10^{17}$ ions/cm$^2$ at $E(^{22}{\rm Ne})$ = 150 keV, and  $0.5 \cdot 10^{17}$ ions/cm$^2$ at $E(^{22}{\rm Ne})$ = 70 keV. Throughout the implantation, the $^{22}$Ne$^+$ beam current density on target was kept at 2 $\mu$A/cm$^{2}$.

\subsection{$\gamma$-ray detection system}

The system used for the detection of the emitted $\gamma$ rays and for the data acquisition system is similar to the one described previously \cite{Schmidt13-PRC}.
The emitted $\gamma$ rays were detected by two high-purity germanium (HPGe) detectors: The first HPGe was positioned at an angle of 55$^{\circ}$ with respect to the incident ion beam direction, its front cap was at 304\,mm distance to the target, and it had 90\% relative efficiency \cite{Gilmore08-Book}. The second HPGe at 90$^{\circ}$ and 152\,mm distance had 60\% relative efficiency (fig. \ref{fig:setup}). The two detectors were surrounded by BGO scintillators used as anti-Compton shields. Each BGO was enclosed in a 2-cm-thick lead shield suppressing the background from environmental radionuclides. Moreover, the BGO detectors were shielded from direct radiation from the target by 7-cm-thick lead collimators. 

Two independent data acquisition systems have been used in parallel: one with a 100-MHz, 14-bit CAEN N1728B digital ADC (providing list mode data) and the other with a standard analog amplification chain and 16384-channel histogramming Ortec 919E ADC unit. 

The absolute full-energy peak efficiency was measured using calibrated $^{137}$Cs, $^{60}$Co, and $^{88}$Y radioactive standards by Physikalisch-Technische Bundesanstalt (PTB, Braunschweig, Germany) with typical activity error of $<$1\%\,(1$\sigma$). The sources had the same geometry (thin plates, with a point-like active spot at the center) as the targets and were placed directly at the target position. 

The efficiency curve was then extended to 10.76 MeV using the $^{27}$Al(p,$\gamma$)$^{28}$Si resonance at 992 keV \cite{Anttila77-NIM, Zijderhand90-NIMA}. This resonance de-excites emitting a 1779-keV $\gamma$ ray (i.e. in the energy range of radioactive sources) and several $\gamma$ rays with higher energy. The beam spot in the $^{27}$Al(p,$\gamma$)$^{28}$Si run was 5\,mm wide, similar to the  $^{22}$Ne(p,$\gamma$)$^{23}$Na beam spot. At the present target-detector distance of 152 (304) mm for the 90$^{\circ}$ (55$^{\circ}$) detector, the beam spot can be safely approximated as a point. 
A relative approach was used to extend the efficiency curve: The efficiency at 1779 keV was derived from the fit of the calibration standard data, and the efficiency at higher energies was then calculated normalizing to the 1779-keV peak, taking into account the well-known branching ratios and angular distributions \cite{Anttila77-NIM}. 

\begin{figure}[h]
\includegraphics[width=\columnwidth]{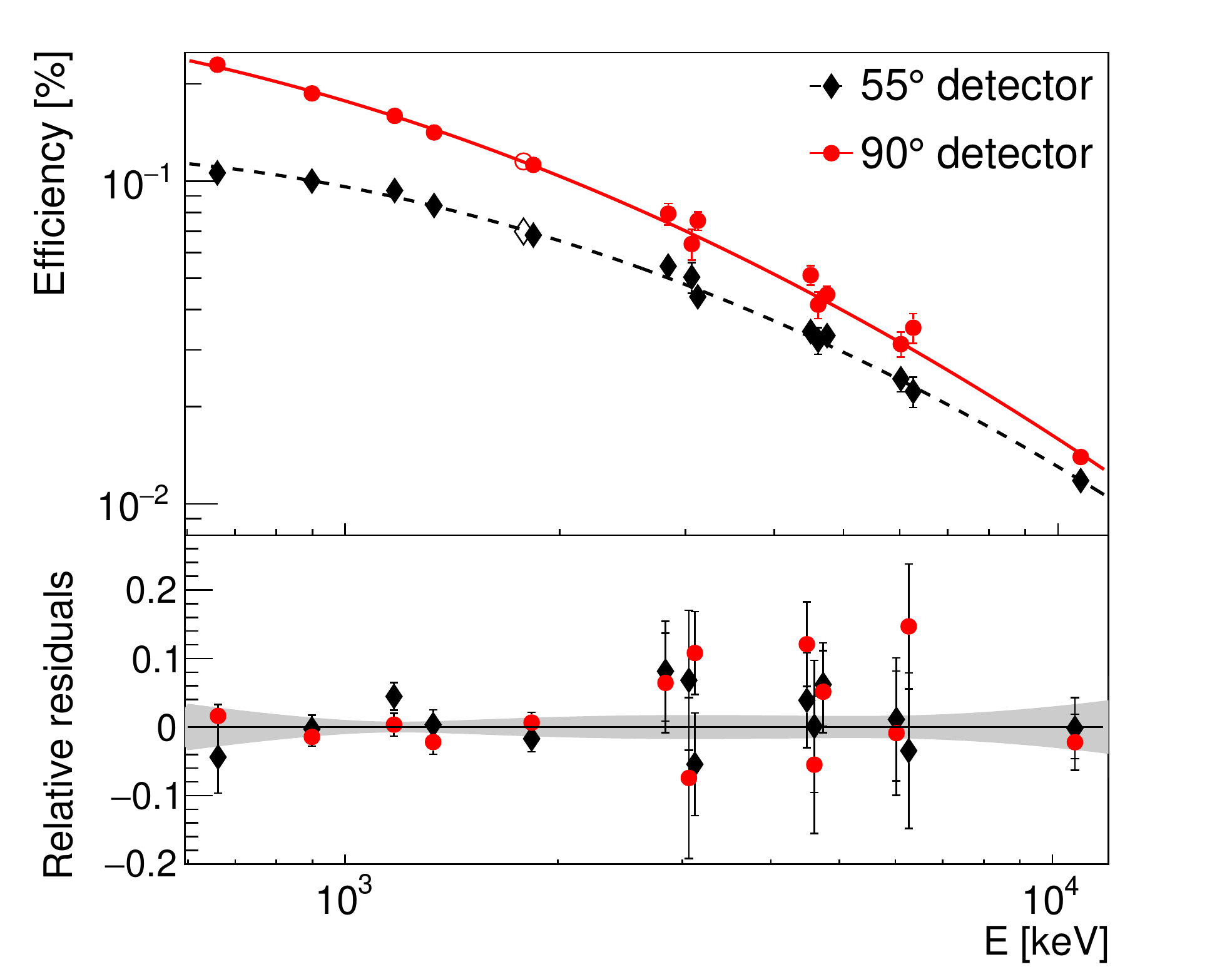}
\caption{\label{fig:Efficiency}Absolute $\gamma$-ray detection efficiency as a function of energy for the two HPGe detectors used. A phenomenological fit function and the fit error are shown.}
\end{figure}

Due to the large distance from source to detector, true coincidence summing effects are well below 1\% for all $\gamma$ lines studied here and have thus been neglected.
The resulting detection efficiency is shown in fig. \ref{fig:Efficiency} for both detectors. The final uncertainty on the detection efficiency is 1-3$\%$, depending on the $\gamma$-ray energy.

\section{Data analysis and results}
\label{sec:Results}

\begin{figure*}[tb]
\includegraphics[width=0.9\textwidth,trim=0 6.1cm 0 0,clip]{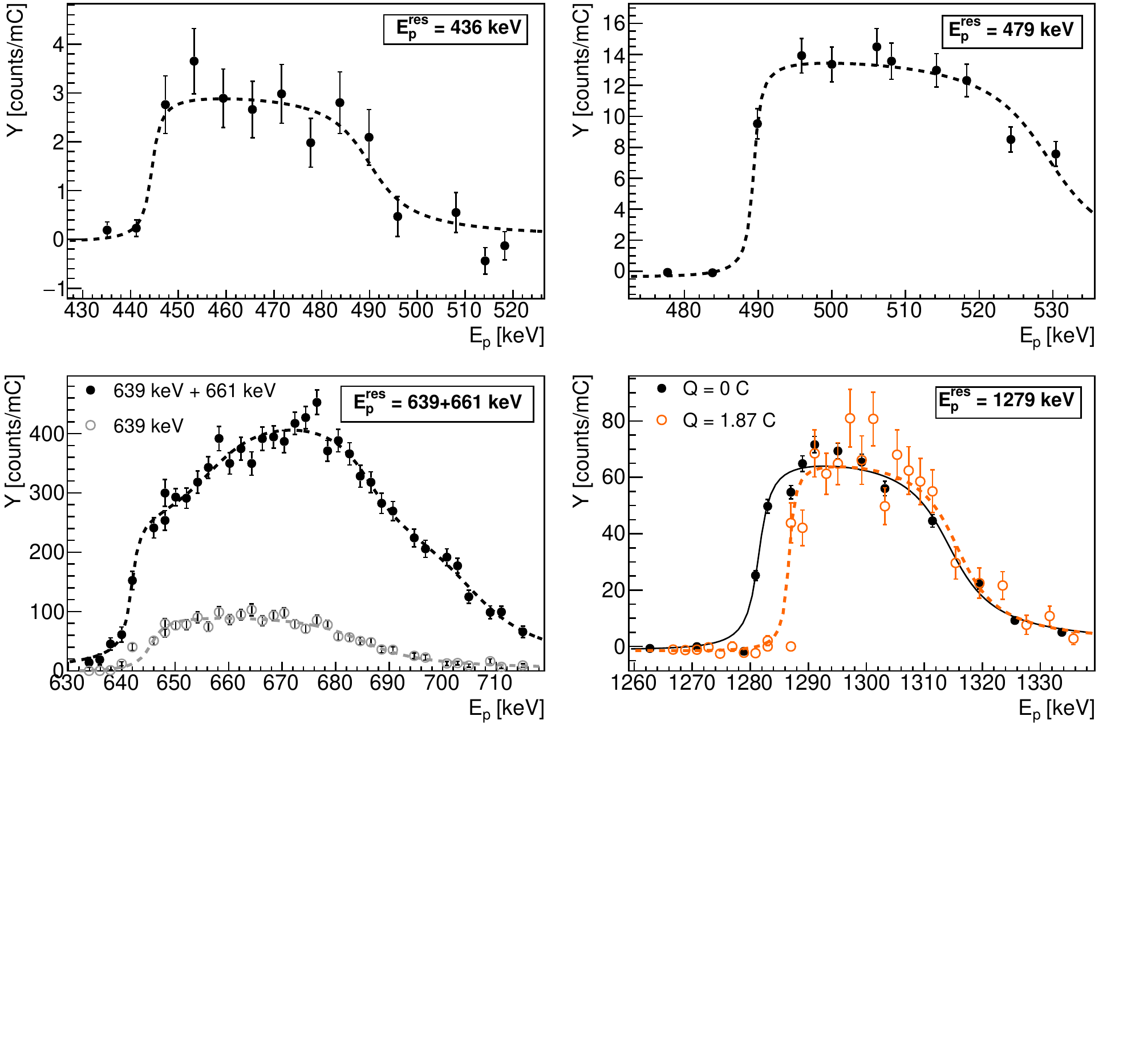}
\caption{\label{fig:scans}Excitation functions of the resonances investigated in the present experiment. The scans are obtained using the following $\gamma$ rays transitions, that can be univocally attributed to the resonance decay (with the exception of the 661-keV resonance): $E_{\rm p}^{\rm res}$ = 436 keV, $E_\gamma$ = 5296. -- $E_{\rm p}^{\rm res}$ = 479 keV, $E_\gamma$ = 6269. -- $E_{\rm p}^{\rm res}$ = 639 keV, $E_\gamma$ = 9403. -- $E_{\rm p}^{\rm res}$ = 639 and 661 keV, $E_\gamma$ = 440. -- $E_{\rm p}^{\rm res}$ = 1279 keV, $E_\gamma$ = 6102\,keV. -- For the 1279-keV resonance, both the scan measured at the beginning (black full circles) and at the end (orange open circles) of the experiment are shown. -- The lines are the fitted Breit-Wigner resonance yield \cite{Iliadis07-Book}, taking beam energy straggling into account.}
\end{figure*}

For each of the five resonances, first the complete excitation function was measured and then a long run was performed at the beam energy of maximum yield. The target stability was monitored by regular scans of the 1279-keV resonance. The general approach is described in section\,\ref{subsec:Approach}. A detailed description of the data analysis for each resonance studied is given in the following sections: Secs.\,\ref{subsec:1279} and \ref{subsec:479} for the two reference resonances, and secs.\,\ref{subsec:436} -- \ref{subsec:661} for the remaining three resonances.

%%%%%%%%%%%%%%%%%%%%%%%%%%%%%%%%%%%%%%%%%%%%%%%%%%%%%%%%%%
\subsection{General approach and target stoichiometry} \label{sec:stoichiometry}\label{subsec:Approach}

\begin{table}[b]
\centering
\begin{ruledtabular}
\begin{tabular}{rc}
$E_{\rm p}^{\rm res}$ & $n$(Ta):$n$($^{22}$Ne) \\ \hline
1279 keV & 8.32 $\pm$ 0.04$_{\rm stat}$ $\pm$ 0.36$_{\rm comm. syst}$ $\pm$ 0.82$_{\omega\gamma}$ \\  
479 keV & 8.41 $\pm$ 0.09$_{\rm stat}$ $\pm$ 0.35$_{\rm comm. syst}$ $\pm$ 0.65$_{\omega\gamma}$  
\\ \hline
Average & 8.37 $\pm$ 0.05$_{\rm stat}$ $\pm$ 0.35$_{\rm comm. syst}$ $\pm$ 0.52$_{\omega\gamma}$ \\
\end{tabular}
\end{ruledtabular}
\caption{Target stoichiometry derived from the resonances at 1279 and 479 keV.}
\label{tab:stoichiometry}
\end{table}

The stoichiometric composition of the target is derived from the effective stopping power $\epsilon$ \cite{Iliadis07-Book}:
\begin{equation}
\epsilon = \frac{\epsilon_{\rm X} n_{\rm X} + \epsilon_{\rm Y} n_{\rm Y} + ...}{n_{\rm X}} = \epsilon_{\rm X} + \frac{n_{\rm Y}}{n_{\rm X}}\epsilon_{\rm Y} + ...
\label{eq:effective-stopp}
\end{equation}
where X is the target nuclide involved in the reaction of interest (here, $^{22}$Ne), Y is a different nuclear species that is not involved in the reaction (here, Ta), $\epsilon_i$ is the stopping power for the element $i$ (in the laboratory system) and $n_i$ is the number of nuclides $i$ per square centimeter. In the approximation of an infinitely-thick target, the effective stopping power is directly related to the measured resonance yield and to the resonance strength $\omega\gamma$:
\begin{equation}
\epsilon = \frac{\lambda^2}{2}\frac{m_p+m_t}{m_t}\frac{\omega\gamma}{Y_{\rm max}}
\label{eq:wg}
\end{equation}
where $\lambda^2/2$ is the de Broglie wavelength (in the center of mass), $m_{\rm p}$ and $m_{\rm t}$ are the masses of the target and projectile and $Y_{\rm max}$ is the experimental yield. 
This approach has then been applied to derive the target stoichiometry Ta:$^{22}$Ne based on the 1279-keV \cite{Keinonen77-PRC} and 479-keV \cite{Longland10-PRC,Kelly15-PRC} resonances (Table~\ref{tab:stoichiometry}). These standards are known with 10$\%$ (7$\%$) precision from the literature, respectively \cite{Keinonen77-PRC, Longland10-PRC,Kelly15-PRC}. 

The systematic uncertainty of the resonance strength is dominated by the uncertainty on the reference value (7\% and 10\% for the two reference resonances at 479 and 1279\,keV, respectively). The other contributions to the systematic uncertainty are charge collection reproducibility (1\%), $\gamma$-ray detection efficiency (1-3\% depending on the $\gamma$-ray energy), energy dependence of the stopping power (2.8\%, based on the 3.7\% SRIM normalization error for Ta \cite{Ziegler10-NIMB} and the relative difference of the stopping powers at the two energies), and data acquisition dead time (0.5\%).

\begin{figure*}[!t]
\includegraphics[width=\textwidth]{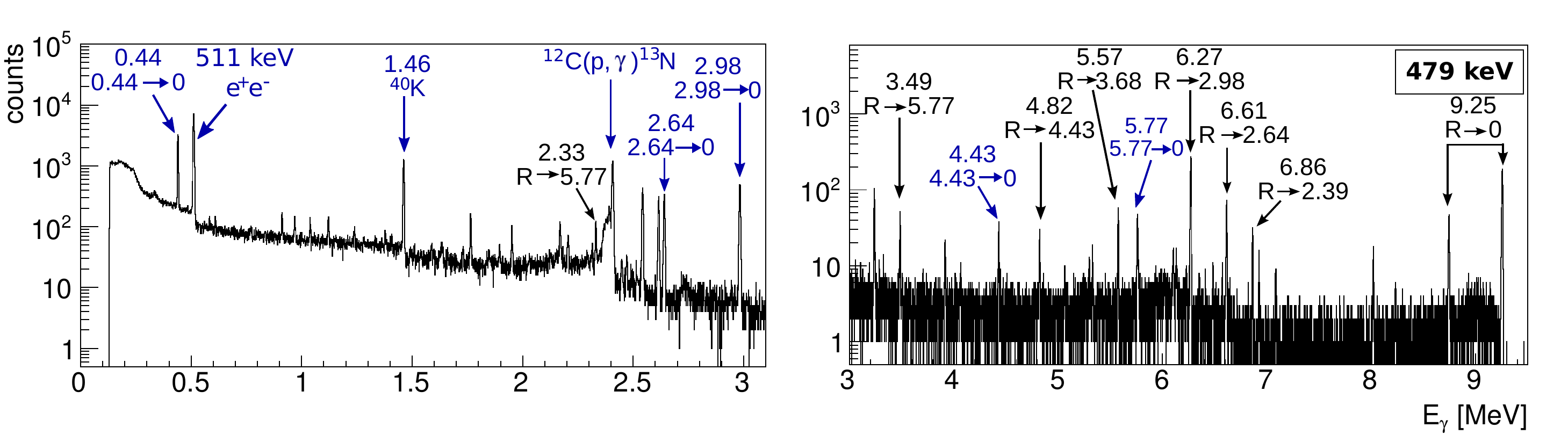}
\includegraphics[width=\textwidth]{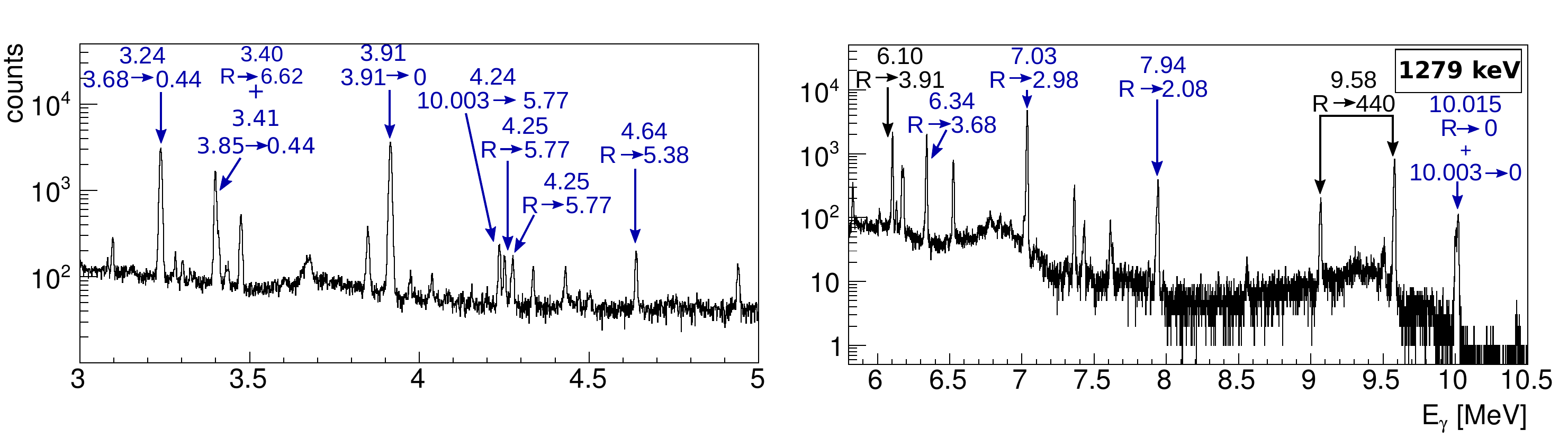}
\caption{\label{fig:479-1279keV_spectra}$\gamma$ ray spectra for the resonances at 479 keV and 1279 keV. The transitions used for the resonance strength determination are shown in black, other $\gamma$ rays in blue.}
\end{figure*}

%%%%%%%%%%%%%%%%%%%%%%%%%%%%%%%%%%%%%%%%%%%%%%%%%%%%%%%%%%
\subsection{1279-keV resonance ($E_{\rm x}$ = 10017 keV)}
\label{subsec:1279}

The 1279-keV resonance is the most intense resonance under study here. Therefore it was also used to monitor the stability of the target over time. 

Fig. \ref{fig:scans} shows two scans of this resonance, one measured at the very beginning of the present experiment (integrated charge $Q$ = 0 C) and the other at the very end ($Q$ = 1.87 C). This one target was used for the whole data taking. Only a shift of the rising edge of the yield profile was observed, which is ascribed to the buildup of contaminants on the target surface.  

In the long run at $E_{\rm p}$ = 1293 keV (fig.\,\ref{fig:479-1279keV_spectra}), the 1279-keV resonance was populated together with the lower-energy resonance at 1264 keV, which according to the literature is ten times weaker \cite{Smit79-NPA}. In order to exclude possible effects due to this secondary resonance, the 1279-keV resonance was analyzed using only two intense primary $\gamma$-rays ($E_\gamma$ = 6102\,keV from the 10017$\rightarrow$3914 transition and $E_\gamma$ = 9575 keV from the 10017$\rightarrow$440 transition) that are well separated from the gammas emitted in the decay of the 1264-keV resonance (fig. \ref{fig:479-1279keV_spectra}). In this way, the scan was unambiguously related to the 1279-keV resonance. 

For the resonance strength analysis, the reaction yield was calculated individually for each transition and for each detector using the following relation:
\begin{equation}
Y_{\rm max} = \frac{N}{\eta \cdot Br \cdot W(\theta) \cdot Q}
\label{eq:yield}
\end{equation}
where N is the net observed peak area (after subtracting continuum background), $\eta$ is the $\gamma$-ray detection efficiency, $B$ is the transition branching ratio, $W(\theta)$ is the angular distribution and $Q$ is the beam integrated charge. The values for $B$ and $W(\theta)$ are taken from the literature \cite{Viitasalo72-ZPhys}.
 
The resulting target stoichiometries derived for each of the two transitions analyzed for both HPGe detectors are mutually consistent, therefore the average value has been adopted for the calculation of the target stoichiometry (table~\ref{tab:stoichiometry}). The contribution to the uncertainty due to the resonance $\omega\gamma$ has been kept separated from the statistical uncertainty and from the systematic uncertainties which are common to the 479-keV resonance, i.e. the detection efficiency (fig.\,\ref{fig:Efficiency}) and the SRIM stopping power \cite{Ziegler10-NIMB}. It is found that the uncertainty on the target composition is dominated by the uncertainty on the resonance strength (table~\ref{tab:stoichiometry}).

%%%%%%%%%%%%%%%%%%%%%%%%%%%%%%%%%%%%%%%%%%%%%%%%%%%%%%%%%%
\subsection{479-keV resonance ($E_{\rm x}$ = 9252 keV)}
\label{subsec:479}

The excitation function and a typical spectrum of the 479-keV resonance are shown in fig. \ref{fig:scans} and \ref{fig:479-1279keV_spectra}, respectively. The 479-keV resonance was well separated in energy from all the other resonances. Moreover, all the gamma transitions reported in the literature for the de-excitation of the resonance were observed. Therefore, the total reaction yield, obtained summing the contributions $N_i$ from all primary transitions $i$ was used for the analysis, making the result independent of the branching ratio found:
\begin{equation}
Y_{\rm max} = \frac{1}{Q} \cdot \sum_{i}{\left(\frac{N_i}{\eta_i} \right)}
\label{eq:tot_yield}
\end{equation}
The angular distribution coefficients for the $\gamma$ rays de-exciting the resonance are not known from the literature. Given that the resonance spin is $1/2$, an isotropical angular distribution for primary and secondary $\gamma$ rays de-exciting this resonance can be safely assumed. This assumption is supported by the fact that the resonance yields measured with the two detectors are perfectly compatible within the error bars. 

The decay branching ratios have also been re-measured. The results are reported in tab. \ref{tab:branching}. 

The strength of the 479-keV resonance can also be calculated using the target stoichiometry derived from the 1279-keV resonance (sec.\ref{subsec:1279}). Inverting equation \ref{eq:wg}, a resonance strength $\omega\gamma(479) =(0.605 \pm 0.006_{\rm stat} \pm 0.062_{\rm syst})$\,eV is found. This value is consistent with the literature strength of $\omega\gamma = (0.583 \pm 0.043) eV$ \cite{Kelly15-PRC} that was derived by depth profiling in aluminium.  

However, the argument can also be turned around. Based on the literature resonance strength \cite{Kelly15-PRC}, the target stoichiometry has also been determined independently of the 1279-keV resonance. Following the approach described in sec.\,\ref{subsec:Approach}, a stoichiometric value is found that is in perfect agreement with the one based on the 1279-keV resonance (table \ref{tab:stoichiometry}). Again, the main contribution to the uncertainty is the error on the resonance strength.

\begin{table}[b!!]
\centering
\begin{ruledtabular}
\begin{tabular}{rrrr}
Final level [keV] & \multicolumn{3}{c}{Resonance energy $E_{\rm p}^{\rm res}$ [keV]} \\ %\cline{2-4}
 & 436 & 479 & 639 \\ \cline{2-4}
Final level [keV] & \multicolumn{3}{c}{Initial level $E_{\rm x}$ [keV]} \\
%\cline{2-4}
 & 9211 & 9252 & 9405 \\ \hline
0 & 1.2 (0.6) & 43.6 (0.9) & 76.0 (0.7) \\
440 & 4.9 (0.5) & & 1.5 (0.1) \\
2391 & 2.0 (0.3) & 3.6 (0.2) &  \\ 
2640 & 6.4 (0.5) & 9.4 (0.3) & 2.9 (0.1) \\
2982 & 22.4 (1.0) & 32.7 (0.6) & 7.8 (0.4)  \\
3678 & 3.1 (0.4) & 4.5 (0.3) & 7.6 (0.2) \\
3848 & 2.0 (0.3) & &  \\
3914 & 30.0 (1.7) & $<$ 0.1 &  \\
4430 & 4.7 (0.4) & 1.9 (0.1) & 3.4 (0.1) \\
5766 &  & 2.0 (0.1) & 0.41 (0.03) \\
5964 & 17.1 (0.8) & &  \\
6195 & 3.4 (0.3) & &  \\
6921 &  & 2.2 (0.1) &  \\
7488 & 2.8 (0.6) & &  \\
7724 &  & & 0.44 (0.03) \\
\end{tabular}
\end{ruledtabular}
\caption{Branching ratios for the $\gamma$-decay of the $E_{\rm p}^{\rm res}$ = 436-, 479-, and 639-keV resonances, corresponding to the $E_{\rm x}$ = 9211-, 9252-, and 9405-keV excited states of $^{23}$Na, respectively. Statistical uncertainties are reported in parentheses. The upper limit is given at 90\% confidence level.}
\label{tab:branching}
\end{table}

For the analysis of the remaining three resonances, it is therefore possible to use the stoichiometry either from the 479- or from the 1279-keV resonance or their average as a standard.

\begin{figure*}[tb]
\includegraphics[width=\textwidth]{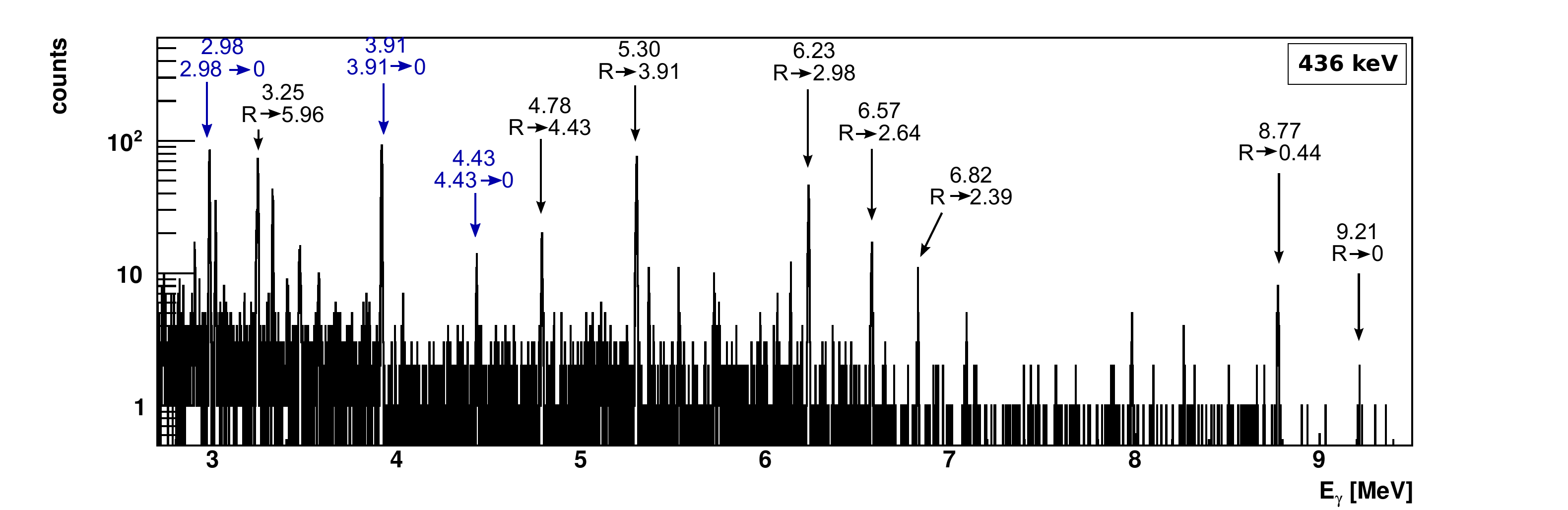}
\includegraphics[width=\textwidth]{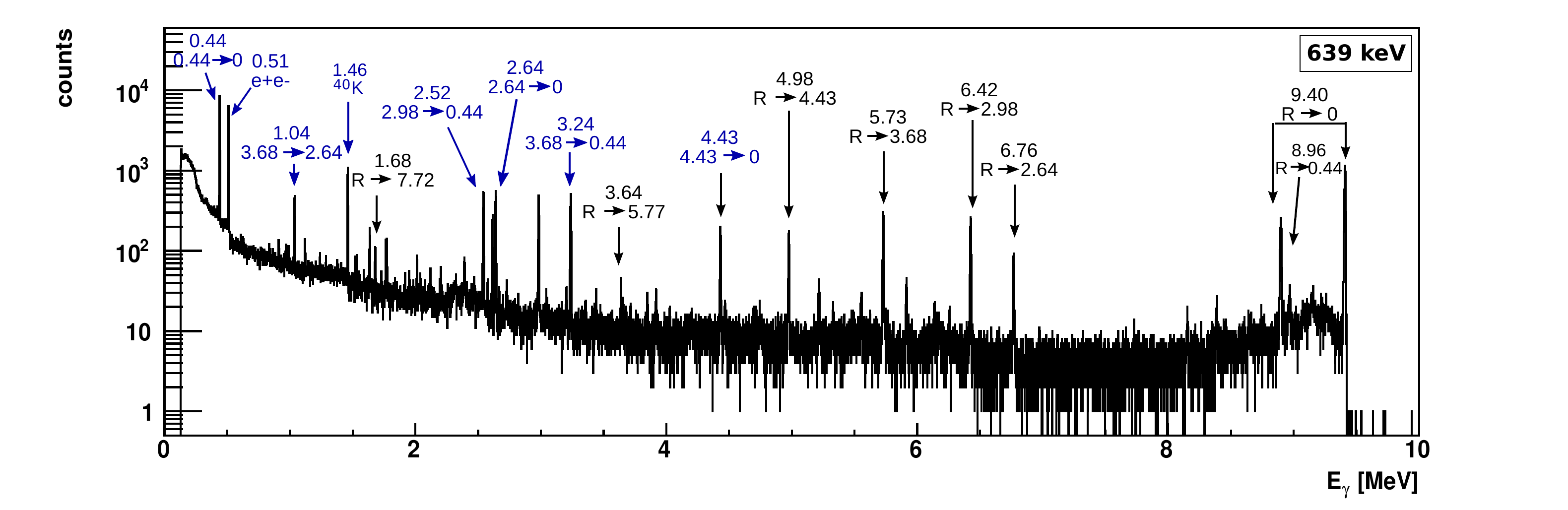}
\includegraphics[width=\textwidth]{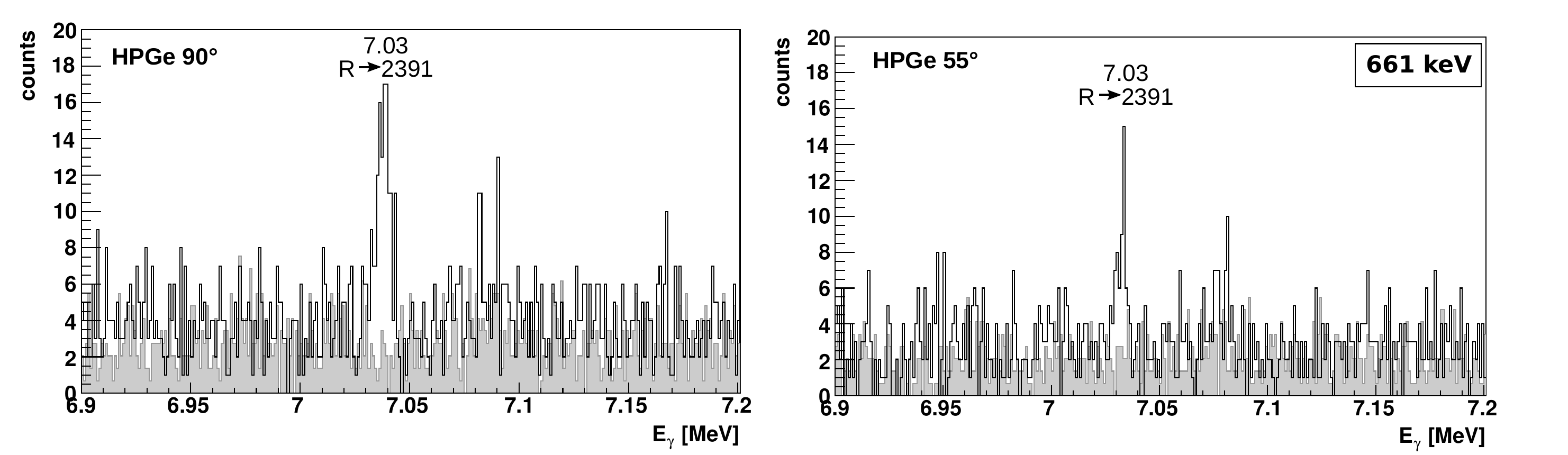}
\caption{\label{fig:spectra}$\gamma$ ray spectra for the resonances at 436 keV, 639 keV and 661 keV. The transitions used for the resonance strength determination are shown in black, other $\gamma$ rays in blue.}
\end{figure*}

%%%%%%%%%%%%%%%%%%%%%%%%%%%%%%%%%%%%%%%%%%%%%%%%%%%%%%%%%%
\subsection{436-keV resonance ($E_{\rm x}$ = 9211 keV)}
\label{subsec:436}

The 436-keV resonance strongly affects the $^{22}$Ne(p,$\gamma$)$^{23}$Na reaction rate at typical temperatures of classical novae explosions. The literature strength value has 23$\%$ uncertainty, $\omega\gamma$ = (0.065 $\pm$ 0.015) eV. The strength of the 436-keV resonance was measured with 30$\%$ uncertainty \cite{Meyer73-NPA} using the 639-keV resonance to convert the relative strengths to an absolute $\omega\gamma$. The value reported in Ref.\,\cite{Meyer73-NPA} was then recalibrated according to the strength of the 1279-keV resonance \cite{Endt90-NPA}. 

A scan of the 436-keV resonance is shown in fig. \ref{fig:scans}. The $\gamma$-ray spectrum from the long run is shown in fig. \ref{fig:spectra}.  
Since all the transitions de-exciting the resonance have been observed, the total reaction yield for each detector has been calculated with eq. \ref{eq:tot_yield}. 
The resonance strength was derived independently for each detector using equation \ref{eq:wg}. Also in this case, no experimental information is available on the angular distribution of emitted $\gamma$-rays. Therefore, calculated correction factors \cite{Iliadis07-Book} of 0.80-1.25, depending on the transition involved and neglecting mixing between different multipolarities, have been used to rescale the 90$^\circ$ data. Subsequently, the average $\omega\gamma$ from both detectors has been adopted to reduce the statistical uncertainty. 

The difference between the 55$^\circ$ yield, which is insensitive to the angular distribution for the present case (where the fourth-order Legendre polynomial can be neglected due to a negligible d-wave contribution), and the combined yield including also the 90$^\circ$ data with their calculated  angular correction is 2\%. This value is included in the systematic error budget for this resonance. The final result (reported in tab. \ref{tab:strengths}) is compatible with the literature strength, but the relative uncertainty has been reduced by a factor of 3. 
Updated branching ratios for the gamma decay of the 9211-keV level have also been derived. The present results (tab. \ref{tab:branching}) are generally in agreement with \cite{Meyer73-NPA}, except for the weakest transitions.

%%%%%%%%%%%%%%%%%%%%%%%%%%%%%%%%%%%%%%%%%%%%%%%%%%%%%%%%%%
\subsection{639-keV resonance ($E_{\rm x}$ = 9405 keV)}
\label{subsec:639}

The literature strength $\omega\gamma$ = (2.8 $\pm$ 0.3) eV was measured and normalized to the strength of the 1279-keV resonance \cite{Keinonen77-PRC}. 
In the present experiment, all the transitions de-exciting the 9405 keV level were observed, therefore the same approach described for the 436-keV resonance has been adopted. 

Also for the 639-keV resonance, no information is available on the $\gamma$ ray angular distribution, but the independent analysis of the spectra recorded with the 55$^{\circ}$ and 90$^{\circ}$ detector gives compatible results. The present resonance strength is consistent with the literature and the uncertainty has been slightly improved.
 
%%%%%%%%%%%%%%%%%%%%%%%%%%%%%%%%%%%%%%%%%%%%%%%%%%%%%%%%%%
\subsection{661-keV resonance ($E_{\rm x}$ = 9426 keV)}
\label{subsec:661}

\begin{table*}[bt]
\centering
\begin{ruledtabular}
\begin{tabular}{cccc}
Detector & Transition & \multicolumn{2}{c}{$\omega\gamma$ [eV]} \\
 & & Relative to 479 keV & Relative to 1279 keV  \\
\cline{2-4}
HPGe 90$^{\circ}$ & R $\rightarrow$ 7488 & 0.056 $\pm$ 0.017$_{stat}$ $\pm$ 0.004$_{syst}$ & 0.055 $\pm$ 0.017$_{syst}$ $\pm$ 0.006$_{syst}$  \\
 & R $\rightarrow$ 7082 & 0.053 $\pm$ 0.016$_{stat}$ $\pm$ 0.005$_{syst}$ & 0.053 $\pm$ 0.016$_{stat}$ $\pm$ 0.006$_{syst}$ \\
 & R $\rightarrow$ 2391 & 0.034 $\pm$ 0.007$_{stat}$ $\pm$ 0.003$_{syst}$ & 0.033 $\pm$ 0.007$_{stat}$ $\pm$ 0.004$_{syst}$ \\
\\ \hline
HPGe 55$^{\circ}$ & R $\rightarrow$ 2391 & 0.024 $\pm$ 0.006$_{stat}$ $\pm$ 0.002$_{syst}$ & 0.024 $\pm$ 0.006$_{stat}$ $\pm$ 0.002$_{syst}$ \\
\end{tabular}
\end{ruledtabular}
\caption{Strength of the 661 keV resonance measured for each HPGe detector and for each $gamma$ transition observed.}
\label{tab:661keV}
\end{table*}

The 661-keV resonance is the weakest resonance investigated in the present experiment. Since the total energy loss of the proton beam in the target is 39 keV, the 661-keV resonance could not be completely separated from the more intense resonance at 639 keV. Fig. \ref{fig:scans} shows the resonance scan obtained with the 440-keV transition from the first excited state to the ground state of $^{23}$Na. This transition is very intense in the $\gamma$ ray spectrum, since it collects the statistics from the decay of most of the higher energy levels. The excitation function shows two rising edges: The first corresponds to the 639-keV resonance, the second (lying on the plateau of the 639-keV resonance) can be attributed to the 661-keV resonance. As a reference, the scan of the 639-keV resonance only, using the 9403-keV $\gamma$-ray that is specific to the 639-keV resonance, is also shown with open circles. 

Due to the very low observed strength of this resonance, only the most intense transitions are observed in the $\gamma$ ray spectrum. Fig. \ref{fig:spectra} shows the 7034-keV transition from the resonance to the 2391-keV excited state as observed in the two germanium detectors, in the long on-resonance run. The gray spectrum is the long run spectrum on the 639 keV ($E_{\rm p}$ = 650 keV), normalized to the integrated charge of the long run on the 661-keV resonance.  

The $\gamma$ transitions observed in each detector are listed in table \ref{tab:661keV}. The resonance strength has been derived from eq. \ref{eq:wg} - \ref{eq:yield} analyzing each transition independently. Branching ratios are adopted from Ref.\,\cite{Meyer73-NPA}, and the angular distribution coefficients are taken from Ref.\,\cite{Bakkum89-NPA}.
Since not all the resonance strengths in table \ref{tab:661keV} are 1$\sigma$ compatible, a conservative approach has been used to derive the final adopted strength: the adopted value is the weighted average of all the values in table \ref{tab:661keV}, and the 1$\sigma$ error bar is expanded to fully cover the maximum and minimum values. 

The resonance strength from the present work is a factor of 11 lower than the literature strength (see tab. \ref{tab:strengths}). The literature $\omega\gamma$ comes from the direct experiment performed by Meyer and Smit in 1973 \cite{Meyer73-NPA}. In that experiment, absolute resonance strengths were derived using the strength of the 639-keV resonance as a reference for normalization. No clear explanation for the discrepancy between the literature and the present strength could be found. It can only be speculated that the Meyer and Smit data were affected by contributions from the stronger resonance at 639\,keV.

\begin{table*}[tb]
\centering
\begin{ruledtabular}
\begin{tabular}{rcccc}
$E_{\rm p}^{\rm res}$ [keV] & \multicolumn{4}{c}{$\omega\gamma$ [eV]} \\
 & Literature & This work & This work & This work  \\
 &  & relative to 479 keV & relative to 1279 & adopted  \\
\\ \hline
436 & 0.065 $\pm$ 0.015 \cite{Meyer73-NPA,Endt90-NPA} & 0.080 $\pm$ 0.002$_{\rm stat}$ $\pm$ 0.007$_{\rm syst}$  & 0.079 $\pm$ 0.002$_{\rm stat}$ $\pm$ 0.008$_{\rm syst}$ & 0.079  $\pm$ 0.006  \\
479 & 0.583 $\pm$ 0.043 \cite{Kelly15-PRC} & & 0.605 $\pm$ 0.006$_{\rm stat}$ $\pm$ 0.062$_{\rm syst}$ & 0.594 $\pm$ 0.038  \\
639 & 2.8 $\pm$ 0.3 \cite{Keinonen77-PRC} & 2.46 $\pm$ 0.02$_{\rm stat}$ $\pm$ 0.21$_{\rm syst}$ & 2.43 $\pm$ 0.02$_{\rm stat}$ $\pm$ 0.25$_{\rm syst}$ & 2.45 $\pm$ 0.18   \\
661 & 0.35 $\pm$ 0.1 \cite{Meyer73-NPA} & 0.032$^{+0.024}_{-0.009}$ & 0.031$^{+0.024}_{-0.009}$ & 0.032$^{+0.024}_{-0.009}$  \\
1279 & 10.5 $\pm$ 1.0 \cite{Keinonen77-PRC} & 11.03 $\pm$ 0.05$_{\rm stat}$ $\pm$ 1.0$_{\rm syst}$ & & 10.8 $\pm$ 0.7 \\
\end{tabular}
\end{ruledtabular}
\caption{Resonance strengths measured in the present work, based on the new normalization for the 479\,keV resonance \cite{Kelly15-PRC}. For the 436, 639, and 661 keV resonances, the value adopted in the last column is the weighted average between the present values relative to the 479 and 1279\,keV resonances (previous two columns). For the 479 and 1279\,keV resonances themselves, the adopted value is formed by the average between the present new value and the previous literature value. See text for details.}
\label{tab:strengths_new}
\end{table*}

%%%%%%%%%%%%%%%%%%%%%%%%%%%%%%%%%%%%%%%%%%%%%%%%%%%%%%%%%%
\section{Discussion}
\label{sec:Discussion}

\begin{figure}[b!]
\includegraphics[width=\columnwidth]{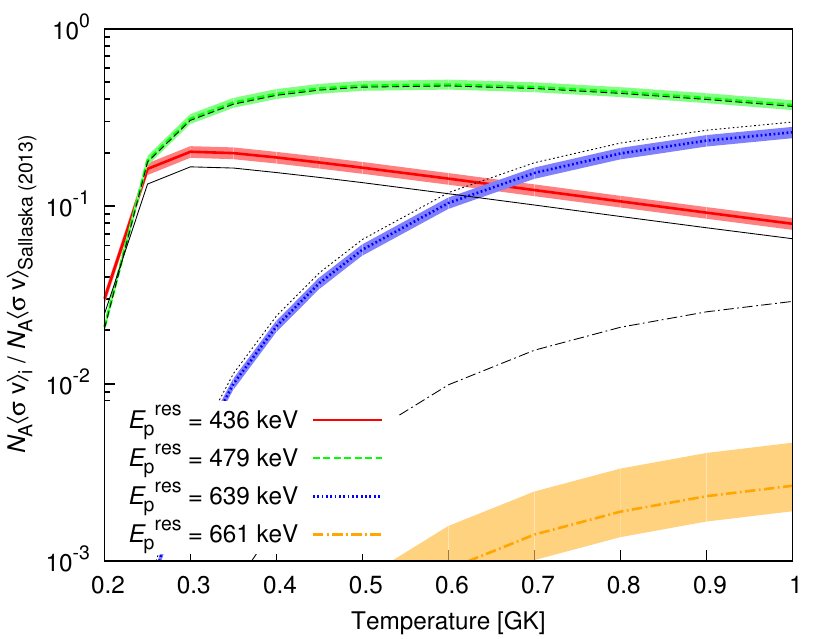}
\caption{\label{fig:rate}The colored lines (colored shaded areas) show the contributions (error bands) $N_{\rm A}\langle\sigma v\rangle_{i}$ of the individual resonances studied to the thermonuclear reaction rate, relative to the Sallaska, Iliadis {\it et al.} \cite{Sallaska13-ApJSS} total reaction rate. For comparison, gray lines with the same line style represent the contributions calculated with the previous literature resonance strengths \cite{Iliadis10-NPA841_31}.}
\end{figure}

The new values for the resonance strength (table~\ref{tab:strengths}) are in agreement with, but more precise than the literature values. The error bars are now on the 8\% level. The two reference resonances at 479 and 1279\,keV now have 7\% systematic uncertainty, when using both the literature data \cite{Longland10-PRC,Keinonen77-PRC} and the new link between them developed in the present experiment.

The only exception is the 661-keV resonance. The new, much lower strength is based on a total of three transitions, one of which was observed in both detectors, so the present value may be considered more reliable than the literature \cite{Meyer73-NPA}.

The impact of each of the resonance strength values $\omega\gamma_i$ determined here on the total thermonuclear reaction rate is quantified by computing the partial thermonuclear reaction rate $N_{\rm A}\langle\sigma v\rangle_{i}$ as a function of temperature $T$
\begin{equation}
N_{\rm A}\langle\sigma v\rangle_{i} = N_{\rm A}\left(\frac{2\pi}{\mu k_{\rm B}T}\right)^\frac{3}{2}\hbar^2\omega\gamma_i \exp\left(-\frac{E^{\rm res}}{k_{\rm B}T}\right)
\end{equation}
for each $\omega\gamma_i$ and dividing it by the total rate from the recent compilation by Sallaska and coworkers \cite{Sallaska13-ApJSS}. In the equation, $N_{\rm A}$ is Avogadro's number, $\mu$ the reduced mass, $k_{\rm B}$ Boltzmann's constant, and $E^{\rm res}$ the resonance energy in the center-of-mass system. 

The review of these partial contributions shows that the resonances at 436, 479, and 639\,keV have an impact of more than 10\% at one given temperature (fig.\,\ref{fig:rate}), whereas the 661- and 1279-keV resonances give a little contribution, $<$1\% of the total reaction rate. The 661-keV resonance was already contributing only negligibly with the previous strength value; with the present, much lower strength it becomes fully negligible. The 1279-keV resonance is located at too high energy, therefore it has no impact at burning temperatures $T_9<1$ and is not plotted in Fig.\,\ref{fig:rate}. 

The present values will contribute to a future new determination of the $^{22}$Ne(p,$\gamma$)$^{23}$Na reaction rate based on low-energy resonances that have recently been studied in the LUNA experiment at Gran Sasso \cite{Cavanna15-PRL,Cavanna15-PhD,Depalo15-PhD}.\\

\section{Summary}
\label{sec:Summary}

The $^{22}$Ne(p,$\gamma$)$^{23}$Na reaction has been investigated in the energy range of interest of classical novae explosions and type Ia supernovae. In total, five resonances have been studied in a direct experiment. 

The strengths of the 436-keV and 639-keV resonances have been determined with 8$\%$ uncertainty, much more precise than in the literature. The strength of the 661-keV resonance has been shown to be 11 times lower than previously believed. The strengths of the 479- and 1279-keV resonances have been determined relative to each other by high-statistics data, and a new value with 7\% precision has been recommended for each of these two resonances.

The new resonance strengths contribute up to 50\% to the total reaction rate and may be used in a future re-evaluation of the $^{22}$Ne(p,$\gamma$)$^{23}$Na thermonuclear reaction rate.

\begin{acknowledgments}
This work was supported in part by DFG (BE4100/2-1), the Helmholtz Association (NAVI, VH-VI-417), EuroGENESIS, and the European Union (FP7-SPIRIT, contract no. 227012).
\end{acknowledgments}

\end{document}